\documentclass[prd,showpacs,preprintnumbers,amsmath,amssymb]{revtex4}

\usepackage{graphicx}
\usepackage{dcolumn}
\usepackage{bm}

\begin{document}

\title{Nonperturbative QED Effective Action at Finite Temperature}
\author{Sang Pyo Kim}\email{sangkim@kunsan.ac.kr}
\affiliation{Department of Physics, Kunsan National University,
Kunsan 573-701, Korea} \affiliation{Asia
Pacific Center for Theoretical Physics, Pohang 790-784, Korea}

\author{Hyun Kyu Lee}\email{hyunkyu@hanyang.ac.kr}
\author{Yongsung Yoon}\email{cem@hanyang.ac.kr}
\affiliation{Department of Physics, Research Institute for Natural
Sciences, Hanyang University, Seoul 133-791, Korea}

\date{\today}

\begin{abstract}
We propose a novel method for the effective action of spinor and
scalar QED at finite temperature in time-dependent electric fields,
where charged pairs evolve in a nonadiabatic way. The imaginary part of the effective
action consists of thermal loops of the Fermi-Dirac or Bose-Einstein
distribution for the initial thermal ensemble, weighted with factors of the Bogoliubov
coefficients for quantum effects. And the real part of the effective action is determined by
the mean number of produced pairs and vacuum polarization at zero temperature.
In the weak-field limit, the mean number of produced pairs is shown twice the
imaginary part. We explicitly find the finite temperature effective
action in a constant electric field.
\end{abstract}

\maketitle

\section{Introduction}

In a strong electromagnetic field the vacuum becomes polarized due
to the interaction of the electromagnetic field with virtual charged
pairs from the Dirac sea. The effective actions in electromagnetic fields
have been continuously investigated since the early work by Sauter, Heisenberg
and Euler, and Weisskopf \cite{Sauter} and later on
the proper-time integral for the
effective action by Schwinger \cite{Schwinger}. The Heisenberg-Euler effective
action exhibits both vacuum polarization and pair production, and has
many physical applications (for a review and references, see Refs.
\cite{Dittrich-Gies,Dunne,RVX}).

The effective actions at zero temperature have been a nontrivial task
for general profiles of electromagnetic fields. As a strong
electric field always creates pairs from the vacuum, the
corresponding effective action contains not only the real part
responsible for vacuum polarization but also the imaginary part for
the decay of vacuum. Thus, the quantum field theory for strong
electric fields should properly handle pair creation from the
vacuum. The effective actions have been found for a pulsed-electric
field of Sauter type in the resolvent method \cite{Dunne-Hall} and
in the evolution operator method \cite{KLY08}, and the effective
action could be found for a spatially localized electric field
\cite{KLY09-2}.

However, the QED effective action at finite temperature in electric field backgrounds
has been an issue of constant interest and controversy, partly because different formalisms
give conflicting results \cite{EPS,Loewe-Rojas,Elmfors-Skagerstam,GKP,Hallin-Liljenberg,Gies99}
 and partly because the thermal effects may
be important to astrophysical objects involving strong
electromagnetic fields. In fact, most methods for finite
temperature field theory may not be directly applied to electric fields
due to pair creation from the vacuum. The one-loop energy momentum tensor of fermions
in an initial thermal ensemble was found in a
constant electric field background \cite{Gavrilov-Gitman}. Recently the closed-time formalism has been employed
to find the QED effective action at finite
temperature in 0+1 dimension \cite{Das-Frenkel}. The enhancement of pair production
by the electric field at finite temperature is also found \cite{Monin-Zayakin}.

The purpose of this paper is two-fold: we first propose a novel method for
the effective action at finite temperature for time-dependent quantum fields and
then find the QED effective action in strong electric field
backgrounds. At zero temperature the effective action is the scattering
amplitude between the out-vacuum and the in-vacuum, which is the
expectation value of the evolution operator with respect to the
in-vacuum \cite{KLY08}. To extend the in- and out-state formalism to finite temperature,
we first express the evolution operator in terms of the Bogoliubov coefficients
and then find the effective action as the expectation value of the evolution operator
with respect to the `thermal
vacuum'. It turns out that the finite temperature effective action is the trace
of the evolution operator weighted with the initial thermal
ensemble of fermions or bosons, which is equivalent to the
`thermal vacuum' expectation value of the evolution operator in thermofield dynamics.
The formalism may be applicable to other time-dependent quantum field,
whose Hamiltonian $H(t)$  evolves in a nonadiabatic way, that is, out of equilibrium,
so that $e^{- \beta H(t)}$ is not the density operator satisfying the Liouville-von Neumann equation \cite{Kim-Lee00}.

We apply the new method to the QED effective action at finite temperature in
time-dependent electric fields. The QED effective action consists of
the zero-temperature part, the part for thermal and vacuum
fluctuations, and the finite temperature part without the electric
field. The logarithm of the Bogoliubov coefficient plays a role of
complex chemical potential in the complex thermal distribution for the effective action.
The real and the imaginary parts of the effective action have an expansion in terms of the
Fermi-Dirac or Bose-Einstein distribution and the chemical
potential. Finally we find the effective action in a constant
electric field and discuss it for the Sauter-type electric field.

The organization of this paper is as follows. In Sec. II, we propose
a new method to find the finite temperature effective action for
time-dependent systems. The effective action is given by the trace
of the evolution operator and the initial density operator, which is
equivalent to the expectation value of the evolution operator
with respect to the thermal vacuum of thermofield dynamics.
In Sec. III, we find the effective action in spinor and scalar QED in
electric fields and then elaborate an expansion scheme in terms of the Fermi-Dirac and
Bose-Einstein distributions. In Sec. IV,  we apply the formalism to
find the effective action in a constant electric field. Finally, we
discuss the controversial issue of thermal effects on
pair-production rate in Sec. V.

\section{Finite Temperature Effective Action}

We consider both spinor and scalar QED with the time-dependent gauge
field $A_{\parallel} (t)$, which generates a constant or
time-dependent electric field. For a pulse-like electric field acting
for a finite period of time, the
ingoing and the outgoing vacua are well-defined at $t_{\rm in} = -
\infty$ and $t_{\rm out} = \infty$, for which we may choose a gauge
$A_{\parallel} (t_{\rm in}) = 0$ such that the ingoing vacuum $\vert
0, t_{\rm in} \rangle$ is nothing but the Minkowski vacuum $\vert 0
\rangle_{\rm M}$. In the case of a constant electric field, we may
use the asymptotic state as in Ref. \cite{KLY08}. The particle and antiparticle
have the momentum ${\bf k}$ and the spin state $\sigma$, whose
annihilation operators are denoted by $a_{{\bf k} \sigma, {\rm in}}$ and $b_{{\bf k} \sigma, {\rm in}}$
at $t_{\rm in} = - \infty$ and $a_{{\bf k} \sigma, {\rm out}}$ and $b_{{\bf k} \sigma, {\rm out}}$
at $t_{\rm out} = \infty$, where $\sigma = \pm 1/2$ for spinor QED and $\sigma = 0$ for scalar QED.
Then, the in- and out-vacua are related through the Bogoliubov
transformations \cite{Kim-Lee07}
\begin{eqnarray}
a_{{{\bf k} \sigma}, {\rm out}} &=& \mu_{{\bf k} \sigma} a_{{{\bf k} \sigma}, {\rm in}} +
\nu^*_{{\bf k} \sigma} b^{\dagger}_{{{\bf k} \sigma}, {\rm in}} = U_{{\bf k} \sigma} a_{{{\bf k} \sigma},
{\rm in}} U^{\dagger}_{{\bf k} \sigma}, \nonumber\\ b_{{{\bf k} \sigma}, {\rm
out}} &=& \mu_{{\bf k} \sigma} b_{{{\bf k} \sigma}, {\rm in}} + \nu^*_{{\bf k} \sigma}
a^{\dagger}_{{{\bf k} \sigma}, {\rm in}} = U_{{\bf k} \sigma} b_{{{\bf k} \sigma}, {\rm in}}
U^{\dagger}_{{\bf k} \sigma}, \label{bog-tran}
\end{eqnarray}
where $U_{{\bf k} \sigma}$ is the evolution operator, whose form in terms
of $\mu_{{\bf k} \sigma}$ and $\nu_{{\bf k} \sigma}$ is explicitly given in Ref. \cite{KLY08}, and the coefficients satisfy the relation
\begin{eqnarray}
|\mu_{{\bf k} \sigma}|^2 + (-1)^{1+ 2 |\sigma|} |\nu_{{\bf k} \sigma}|^2 = 1. \label{bog rel}
\end{eqnarray}

In the in- and out-state formalism elaborated in Ref. \cite{KLY08}, the in- and out-vacua are annihilated by
$a_{{{\bf k} \sigma}, {\rm in/out}}$ and $b_{{{\bf k} \sigma}, {\rm in/out}}$.
In fact, the in- and out-vacua are the tensor product of the
zero-number states for all ${\bf k}$ and $\sigma$. The evolution operator
transforms the in-vacuum to the out-vacuum as $\vert 0, {\rm out} \rangle = U \vert 0, {\rm in} \rangle$,
where $U$ is also the tensor product of each $U_{{\bf k} \sigma}$, that is, $U =
\prod_{{\bf k} \sigma} U_{{\bf k} \sigma}$. The zero-temperature effective action
per unit volume and per unit time is obtained from the scattering amplitude \cite{KLY08}
\begin{eqnarray}
e^{i \int d^3 x dt {\cal L}_{\rm eff}} = \langle 0, {\rm out} \vert 0,
{\rm in} \rangle = \langle 0, {\rm in} \vert U^{\dagger} \vert 0,
{\rm in} \rangle. \label{zero eff}
\end{eqnarray}
Now we extend the zero-temperature effective action to the finite-temperature one for the system with the
initial density operator,
\begin{eqnarray}
\rho_{\rm in} = \prod_{{\bf k} \sigma}  \Bigl[e^{- n_{{\bf k} \sigma} \beta E({\bf k}, \sigma)} \vert
n_{{\bf k} \sigma}, {\rm in} \rangle \langle n_{{\bf k} \sigma}, {\rm in} \vert \Bigr],
\end{eqnarray}
where $\beta = 1/k_{\rm
B} T$, $k_{\rm B}$ being the Boltzmann constant, and $E({\bf k}, \sigma)$ is the energy
for massive charged particles and antiparticles.\footnote{The unit system of $c = \hbar = k_{\rm B} =1$ is used,
where $qE/m^2$ and $\beta m$ are dimensionless in QED.}

In finite temperature field theories for static systems, one employs either the partition
function $Z (\beta) = {\rm Tr} (\rho)$ or the thermal
expectation value $ \langle O \rangle_{\beta} = {\rm Tr} (O
\rho)/{\rm Tr} (\rho)$, which is equivalent to
the `thermal vacuum' expectation value, $\langle O \rangle_{\beta} =
\langle 0, \beta, {\rm in} \vert O \vert 0, \beta, {\rm in}
\rangle$ \cite{Takahashi,Umezawa,Das}.  However, finite temperature field theories
should be modified for time-dependent systems since they nonadiabatically evolve the initial states.
For such time-dependent quantum fields we propose the finite-temperature effective action
\begin{eqnarray}
e^{i \int d^3 x dt {\cal L}_{\rm eff} (T)} =  \frac{{\rm Tr}
(U^{\dagger} \rho_{\rm in})}{{\rm Tr} (\rho_{\rm in})}. \label{T
eff}
\end{eqnarray}
In fact, the effective action (\ref{T eff}) is equivalent to $\langle 0, \beta, {\rm
in} \vert U^{\dagger} \vert 0, \beta, {\rm in} \rangle $ for the `thermal vacuum' \cite{Takahashi,Umezawa}
\begin{eqnarray}
\vert 0, \beta, {\rm in} \rangle \equiv \frac{1}{Z^{-1/2}_{\rm in}}
\prod_{{\bf k} \sigma} \Bigl[ \sum_{n_{{\bf k} \sigma}}  e^{- n_{{\bf k} \sigma} \beta E({\bf k}, \sigma)/2}
\vert n_{{\bf k} \sigma}, {\rm in} \rangle \otimes \vert \tilde{n}_{{\bf k} \sigma},
{\rm in} \rangle \Bigr],
\end{eqnarray}
where $\vert \tilde{n}_{{\bf k} \sigma}, {\rm in} \rangle$ denotes the state
for a noninteracting fictitious system of the
extended Hilbert space. In fact, Eq. (\ref{T eff}) has the correct zero-temperature
limit (\ref{zero eff}). The effective action (\ref{T eff}) may be applied to other time-dependent quantum fields
as well as QED, which evolve in a nonadiabatic way.

\section{QED Effective Action at $T$}

We now advance a method to compute the QED effective action (\ref{T
eff}) in electric fields. Evaluating Eq. (\ref{T eff}), we obtain
the effective action at finite temperature per unit volume and per unit time,
\begin{eqnarray}
{\cal L}_{\rm eff} (T, E) = (-1)^{2 |\sigma|} i \sum_{{\bf k} \sigma} \Bigl[ - \beta z_{{\bf k} \sigma} + \ln (1 + (-1)^{1+ 2 |\sigma|}
e^{- \beta (\omega_{\bf k} - z_{{\bf k} \sigma})})  - \ln (1+
(-1)^{1+ 2 |\sigma|} e^{- \beta \omega_{\bf k}}) \Bigr], \label{QEDeff}
\end{eqnarray}
where $\omega_{\bf k} = \sqrt{m^2 + {\bf k}_{\perp}^2 + (k_{\parallel} + qA_{\parallel})^2}$ and
\begin{eqnarray}
\frac{1}{\mu_{{\bf k} \sigma}^*} = e^{\beta z_{{\bf k} \sigma}}, \quad (z_{{\bf k} \sigma} =
z_r ({{\bf k}, \sigma}) + i z_i ({{\bf k}, \sigma})). \label{ch pot}
\end{eqnarray}
The summation is over all possible states such as momenta and spin
states. Each term in Eq. (\ref{QEDeff}) has the
following interpretation: the first term is the effective
action ${\cal L}_{\rm eff} (T=0, E)$ at zero temperature, the second term
is the combined effect of thermal and quantum fluctuations, while the
last term is the subtraction of the effective action (potential
energy) ${\cal L}_{\rm eff} (T, E=0)$ at finite temperature without
the electric field. From now on we subtract the zero-temperature
part from the effective action and let
\begin{eqnarray}
\Delta {\cal L}_{\rm eff} (T, E) = {\cal L}_{\rm eff} (T, E) - {\cal
L}_{\rm eff} (0, E). \label{eff dif}
\end{eqnarray}
Note that $z_{{\bf k} \sigma}(E)$, which depends on the electric field $E$
and $z_{{\bf k} \sigma}(0) = 0$, plays a role of complex chemical potential,
as will be explained below.

Further, we elaborate an expansion scheme for the effective action
in terms of the Fermi-Dirac or Bose-Einstein distribution and
$z_{{\bf k} \sigma}$. First, the imaginary part of the effective action
(\ref{eff dif}) can be expanded as
\begin{eqnarray}
{\rm Im} (\Delta {\cal L}_{\rm eff}) =  (-1)^{1+ 2 |\sigma|} \frac{1}{2} \sum_{{\bf k} \sigma}
\sum_{j=1}^{\infty} \frac{[(-1)^{2 |\sigma|} n_{\rm F/B}({\bf k})]^j}{j} [ (e^{\beta
z_{{\bf k} \sigma}} - 1)^j + (e^{\beta z^*_{{\bf k} \sigma}} - 1)^j ], \label{im}
\end{eqnarray}
where $n_{\rm F/B} ({\bf k})$ denotes either the Fermi-Dirac distribution
$n_{F}({\bf k}) = 1/(e^{\beta \omega_{\bf k}} +1)$ for spinor QED or
$n_{B} ({\bf k}) = 1/(e^{\beta \omega_{\bf k}} -1)$ for scalar QED.
Second, the real part of the effective action (\ref{eff dif}) is given by
\begin{eqnarray}
{\rm Re}(\Delta {\cal L}_{\rm eff}) = (-1)^{2 |\sigma|} \sum_{{\bf k} \sigma}
\sum_{j=1}^{\infty} \frac{[(-1)^{2 |\sigma|} e^{- \beta (\omega_{\bf k} -
z_r({\bf k}, \sigma))}]^j }{j} \sin(j \beta z_i ({\bf k}, \sigma)). \label{re}
\end{eqnarray}
The thermal factors in Eqs. (\ref{im}) and (\ref{re}) correspond to thermal loops in the diagrammatic representation, which are weighted with factors from quantum fluctuations

Now, we give physical interpretations for the effective action. In
the weak-field limit $(qE \ll m^2)$ where $\beta z_i ({\bf k}) \ll
1$, the real part (\ref{re}) approximately is
\begin{eqnarray}
{\rm Re} (\Delta {\cal L}_{\rm eff}) \approx \sum_{{\bf k} \sigma}
\frac{\beta z_i ({\bf k}, \sigma)}{e^{\beta (\omega_{\bf k} - z_r({\bf k}, \sigma))
} + (-1)^{1+ 2 |\sigma|} }, \label{re appr}
\end{eqnarray}
while the imaginary part (\ref{im}) approximately leads to
\begin{eqnarray}
2 {\rm Im} (\Delta {\cal L}_{\rm eff}) \approx (-1)^{2 |\sigma|} \sum_{{\bf k} \sigma}
|\nu_{{\bf k} \sigma}|^2 n_{F/B} ({\bf k}). \label{therm eff}
\end{eqnarray}
Thus, the imaginary part may be regarded as the pair-production rate
due to thermal and quantum effects. The thermal effects suppress the fermion
pair production due to the Pauli blocking but enhance the boson pair
production due to the Bose-Einstein condensation, as expected. In
Ref. \cite{Kim-Lee07}, the mean number of produced pairs with a given
momentum ${\bf k}$ at $T$ is given by $\bar{\cal N}^{\rm sp} (T)
= \sum_{{\bf k} \sigma} |\nu_{{\bf k} \sigma}|^2 \tanh (\beta \omega_{\bf k}/2)$ for spinor QED and
$\bar{\cal N}^{\rm sc} (T) = \sum_{{\bf k} \sigma} |\nu_{{\bf k} \sigma}|^2 \coth (\beta \omega_{\bf
k}/2)$ for scalar QED. So the mean number, $\Delta \bar{\cal N} =
(\bar{\cal N}(T) - \bar{\cal N} (0))/2$, of one species of particle
or antiparticle due to thermal effects approximately satisfies the
relation between the mean number and the imaginary part:
\begin{eqnarray}
\Delta \bar{\cal N} = \sum_{{\bf k} \sigma} |\nu_{{\bf k} \sigma}|^2 n_{F/B} ({\bf k}) \approx 2
{\rm Im} (\Delta  {\cal L}_{\rm eff}). \label{T-gen rel}
\end{eqnarray}
The relation between the mean number of produced pairs and twice of the imaginary
part also holds at $T= 0$ in the weak-field limit \cite{KLY08}.

A few comments are in order. The series of the real part (\ref{re})
may be summed as \cite{PBM}
\begin{eqnarray}
{\rm Re}(\Delta {\cal L}_{\rm eff}) = \sum_{{\bf k} \sigma} \arctan
\Bigl[\frac{\sin(\beta z_i ({\bf k}) )}{e^{\beta (\omega_{\bf k} -
z_r({\bf k})) } + (-1)^{1+ 2 |\sigma|} \cos(\beta z_i ({\bf k}))}\Bigr]. \label{re 2}
\end{eqnarray}
Using ${\cal L}_{\rm eff} (0, E) = (-1)^{1+ 2 |\sigma|} i \sum_{{\bf k} \sigma} \beta z_{{\bf
k} \sigma}$ from Eq. (\ref{QEDeff}) and the Bogoliubov relation (\ref{bog rel}), we have the real and imaginary parts
\begin{eqnarray}
\beta z_r ({\bf k}, \sigma) &=& (-1)^{1+ 2 |\sigma|} {\rm Im} ({\cal L}_{\rm eff} (0, E)) = -
\frac{1+ 2 |\sigma|}{2} \ln (1 + (-1)^{2 |\sigma|} |\nu_{{\bf k} \sigma}|^2), \nonumber\\
\beta z_i ({\bf k}, \sigma) &=& (-1)^{2 |\sigma|} {\rm Re} ({\cal L}_{\rm eff} (0, E)).
\end{eqnarray}
Then the effective action (\ref{re 2}) at finite temperature can be
written in terms of the mean number, the vacuum
polarization, and the thermal distribution as
\begin{eqnarray}
{\rm Re}(\Delta {\cal L}_{\rm eff}) = (-1)^{2 |\sigma|} \sum_{{\bf k} \sigma} \arctan
\Bigl[\frac{\sin({\rm Re} ({\cal L}_{\rm eff} (0, E)) )}{e^{\beta
\omega_{\bf k}} (1 + (-1)^{2 |\sigma|} |\nu_{{\bf k} \sigma}|^2)^{\frac{ 1+ 2 |\sigma|}{2}} + (-1)^{1+ 2 |\sigma|} \cos({\rm Re}
({\cal L}_{\rm eff} (0, E)))}\Bigr]. \label{re 3}
\end{eqnarray}
Another interesting observation is that the mid-term in Eq. (\ref{QEDeff}),
\begin{eqnarray}
W_{\rm eff} (T, E) = (-1)^{1+ 2 |\sigma|} i \sum_{{\bf k} \sigma} \ln (1 + (-1)^{1+ 2 |\sigma|} e^{- \beta
(\omega_{\bf k} - z_{{\bf k} \sigma})}), \label{pot en}
\end{eqnarray}
is reminiscent of the potential energy \cite{Kapusta} and carries both thermal and quantum effects.
Equation (\ref{pot en}) suggests $z_{{\bf k} \sigma}$ as the chemical potential and the variation with respect to $z_{{\bf k} \sigma}$ yields the Fermi-Dirac or Bose-Einstein distribution.

\section{Applications}

In this section we find the QED effective action  in
a constant electric field and discuss a Sauter-type electric field, $E(t)
= E_0 {\rm sech}^2(t/\tau)$ in Ref. \cite{KLY08}. In the time-dependent gauge, $A_{\parallel}
(t) = -Et$ for the constant electric field and $A_{\parallel} (t) = -E_0 \tau (1
+ \tanh(t/\tau))$ for the Sauter-type electric field, the energy of charged
particles explicitly depend on time.
We will take the weak-field limit $(qE \ll m^2)$, where the real part (\ref{re appr}) and the imaginary part (\ref{therm eff}) of the approximate effective action can be worked out for the constant electric field and in principle for the Sauter-type electric field.

In the constant electric field, the state along the direction of the electric
field is asymptotically determined, whose momentum integral
gives a factor $qE/(2 \pi)$ \cite{KLY08}. Using the mean number of produced
pairs,
\begin{eqnarray}
|\nu_{{\bf k} \sigma}|^2 = e^{- \pi \frac{m^2 + {\bf k}_{\perp}^2}{qE}},
\end{eqnarray}
which is independent of the spin states, the imaginary part (\ref{therm eff})  is given by
\begin{eqnarray}
{\rm Im} (\Delta {\cal L}_{\rm eff} (T, E))  & \approx& \frac{{1+ 2 |\sigma|}}{2}
\Bigl( \frac{qE}{2 \pi} \Bigr)^2 e^{-\frac{\pi m^2}{qE}}  \sum_{n = 0}^{\infty} (-1)^{2 |\sigma|(n+1) } \frac{m^2 e^{- \beta m(n+1)}}{2 \pi m^2
+ \beta m qE (n+1)} \nonumber\\ && \times \Bigl[1+ \frac{\beta m (qE)^2}{(2 \pi m^2
+ \beta m qE (n+1))^2} - \frac{3 \beta m (qE)^3}{(2 \pi m^2
+ \beta m qE (n+1))^3} +  \cdots \Bigr].
\label{im-conE}
\end{eqnarray}
The factor in front of the summation is the leading term of the imaginary part at zero temperature. Further, in the low-temperature limit $(\beta m \gg 1)$, the leading term of Eq. (\ref{im-conE}) is
\begin{eqnarray}
{\rm Im} (\Delta  {\cal L}_{\rm eff} (T, E))  & \approx& \frac{{1+ 2 |\sigma|}}{2}
\Bigl(\frac{qE}{2 \pi} \Bigr)^2 e^{-\frac{\pi m^2}{qE}} \Bigl[(-1)^{2 |\sigma|} \frac{m^2}{qE} \frac{e^{- \beta m}}{\beta m+ \frac{2 \pi m^2}{qE}}  \Bigr].
\end{eqnarray}
In the special case of thermal effect dominance, neglecting all terms of $m/\beta qE$, the first series in Eq.  (\ref{im-conE}) approximately leads to
\begin{eqnarray}
{\rm Im} (\Delta  {\cal L}_{\rm eff} (T, E))  & \approx& \frac{{1+ 2 |\sigma|}}{2} \Bigl(\frac{qE}{2 \pi} \Bigr)^2 e^{-\frac{\pi m^2}{qE}}  \Bigl[- \frac{m^2}{\beta m}  \ln (1+ (-1)^{1+ 2 |\sigma|}e^{- \beta m})
\Bigr]. \label{im-conE2}
\end{eqnarray}

Similarly, using ${\rm Re} ({\cal L}_{\rm eff} (0, E))$ in Ref. \cite{KLY08},
the real part (\ref{re appr}), for instance, of spinor QED is given by
\begin{eqnarray}
{\rm Re} (\Delta  {\cal L}^{\rm sp}_{\rm eff} (T, E))  & \approx& -
\frac{qE}{2 \pi} \frac{m^2}{2 \pi}  \sum_{n = 0}^{\infty} (-1)^n \sum_{l = 2}^{\infty}
\frac{2^{4l-2} |B_{2l}| }{(2l)!} \Bigl(\frac{qE}{2 \pi} \Bigr)^{2l-1} \frac{1}{m^{4l-2}}
\nonumber\\&& \times \Bigl[ e^{- \beta m (n+1)} \Bigl( \Psi (1, 3 - 2l, \alpha) + \frac{\beta m (n+1)}{4} \Psi (3, 5 - 2l, \alpha) - \frac{3 \beta m (n+1) }{8} \Psi (4, 6 - 2l, \alpha) + \cdots \Bigr) \nonumber\\&&
- e^{- \beta m n} \Bigl( \Psi (1, 3 - 2l, \gamma) + \frac{\beta m n}{4} \Psi (3, 5 - 2l, \gamma) - \frac{3 \beta m n }{8} \Psi (4, 6 - 2l, \gamma) + \cdots \Bigr) + \cdots \Bigr],
\label{re-conE}
\end{eqnarray}
where $B_{2l}$ is the Bernoulli number, $\Psi$ denotes the second confluent hypergeometric function \cite{PBM-Econ}, and
\begin{eqnarray}
\alpha = \frac{\beta m (n+1)}{2}, \quad \gamma = \frac{\beta m (n+1)}{2} + \frac{\pi m^2}{qE}.
\end{eqnarray}
In the low-temperature limit $(\beta m \gg 1)$, the series of $l= 2$ in Eq. (\ref{re-conE}) leads to
\begin{eqnarray}
{\rm Re} (\Delta  {\cal L}^{\rm sp}_{\rm eff} (T, E))  & \approx& -
\frac{(2 \pi)^2}{45 m^4} \Bigl( \frac{qE}{2 \pi} \Bigr)^4 \Bigl[ (-1)
 \frac{3}{\beta m} \ln (1 + e^{- \beta m}) \Bigr]. \label{re-conE2}
\end{eqnarray}
Here the factor in front of the square bracket is the real part at zero temperature. The real part of effective action in scalar QED may be found in a similar way.

Finally, we discuss the Sauter-type electric field. The charged particle
has the free energy $\omega_{{\bf k}, {\rm in}} = \sqrt{m^2 + {\bf k}^2}$
before the onset of the electric field while it has $\omega_{{\bf k}, {\rm out}}
= \sqrt{m^2 + {\bf k}^2_{\perp} + (k_z - 2 qE_0 \tau)^2}$ after
the completion of the interaction. At zero temperature, the mean number
of produced pairs, Eqs. (68) and (83), and the vacuum polarization, Eqs. (66) and (80)
of Ref. \cite{KLY08},  which depend on
$\omega_{{\bf k}, {\rm in}}$, $\omega_{{\bf k}, {\rm out}}$, and $\lambda = \sqrt{(qE_0 \tau^2)^2 - (2|\sigma| -1)^2/4}$, lead to the effective action, Eqs. (\ref{re appr}) and (\ref{therm eff}).
To find analytical expressions for the effective action
would be more complicated than the constant electric field, which will be addressed elsewhere.

\section{Conclusion}

In this paper we have advanced a new method for the finite-temperature effective action
for time-dependent quantum fields and have studied the one-loop effective action
of spinor and scalar QED at finite temperature in a constant or time-dependent electric fields.
External electric fields make the vacuum unstable against pair production,
which is a consequence of the out-vacuum differing from the in-vacuum. The instability
enforces a careful application of the finite-temperature field theory to
time-dependent quantum fields. The finite-temperature effective
action (\ref{T eff}) is given by the trace of the initial thermal ensemble evolved
by the time-evolution operator, which is equivalent to
the thermal vacuum expectation value of the evolution operator in thermofield dynamics.

The imaginary part (\ref{im}) of the effective action exhibits
factorization into thermal factors and quantum factors,
which correspond to thermal loops in the diagrammatic representation
with vertices of the external electric field. In the weak-field limit $(qE \ll m^2)$,
twice of the imaginary part is the mean number of produced pairs,
as shown in Eq. (\ref{therm eff}). However, the thermal
and quantum effects are intertwined in the real part of the effective action,
Eqs. (\ref{re}), (\ref{re 2}), and (\ref{re 3}).
In fact, the finite-temperature effective action (\ref{re 3}) is
determined by the vacuum polarization at zero temperature,
the mean number of produced pairs, and thermal distribution.
In the weak-field and lower-temperature limits, the leading factors of the real and imaginary parts, (\ref{im-conE2}) and (\ref{re-conE2}), of the effective action in a constant electric field
are proportional to those at zero temperature and the potential energy for the
rest mass in spinor and scalar QED.

Our results show many interesting aspects. First, the
imaginary part (\ref{im}) of the thermal contribution
does not vanish for any non-zero electric field, which implies thermal effects
on pair production and thus may resolve the controversial issue of thermal effects on pair
production: thermal effects are shown to exist in Refs.
\cite{Loewe-Rojas,Hallin-Liljenberg}, while in Refs.
\cite{Elmfors-Skagerstam} no thermal effects are found.
Further, in the weak-field limit for small pair production,
twice of the imaginary parts (\ref{T-gen rel}) are the
pair-production rate at $T$, which was shown in Ref. \cite{Kim-Lee07}.
Though our in- and out-state formalism differs from the imaginary-time
formalism, the imaginary part (\ref{therm eff})
in the weak-field limit is the pair-production rate times the Fermi-Dirac or
Bose-Einstein distribution, which may correspond to
two-loop dominance in Ref. \cite{Gies99}. Second, the Bogoliubov coefficient (\ref{ch pot}), which is
responsible for vacuum polarization at $T =0$, plays a role of
chemical potential in the effective action (\ref{QEDeff}) and
in the potential energy (\ref{pot en}) at $T$.
In fact, the variation of the effective action with respect to the
chemical potential yields the Fermi-Dirac or the Bose-Einstein distribution.

\acknowledgments

The authors would like to thank Holger Gies for useful information
and comments, Ismail Zahed for useful discussions, and Sergey Gavrilov for useful comments.
S.~P.~K. would like to thank W-Y. Pauchy Hwang for the warm hospitality
at National Taiwan University, where part of this paper was written.
The work of S.~P.~K. was supported by the Korea
Research Foundation (KRF) Grant funded by the Korea Ministry of
Education, Science and Technology (2009-0075-773) and the work of
H.~K.~L. was supported by the World Class University Program
(R33-2008-000-10087-0) of the Korea Ministry of Education, Science
and Technology.

\end{document}